# Evolution of unidirectional random waves in shallow water (the Korteweg - de Vries model)

Anna Kokorina[1,2)] and Efim Pelinovsky[1,3)]

[1)]Laboratory of Hydrophysics and Nonlinear Acoustics, Institute of Applied Physics, Nizhny Novgorod, Russia; email: enpeli@hydro.appl.sci-nnov.ru
[2)]Applied Mathematics Department, Nizhny Novgorod State Technical University, Nizhny Novgorod, Russia
[3)]IRPHE – Technopole de Chateau-Gombert, Marseille, France

**Abstract:**

Numerical simulations of the unidirectional random waves are performed within the Korteweg -de Vries equation to investigate the statistical properties of surface gravity waves in shallow water. Nonlinear evolution shows the relaxation of the initial state to the quasi-stationary state differs from the Gaussian distribution. Significant nonlinear effects lead to the asymmetry in the wave field with bigger crests amplitudes and increasing of large wave contribution to the total distribution, what gives the rise of the amplitude probability, exceeded the Rayleigh distribution. The spectrum shifts in low frequencies with the almost uniform distribution. The obtained results of the nonlinear evolution of shallow-water waves are compared with known properties of deep-water waves in the framework of the nonlinear Schrodinger equation.

## I. INTRODUCTION

Due to the dispersion of the surface water waves, each individual sine wave travels with a frequency dependent velocity, and they propagate along different directions. Due to the nonlinearity of the water waves, sine waves interact with each other generating new spectral components. As a result, the wave field gives rise to an irregular sea surface that is constantly changing with time. To explain the statistical characteristics of the random wave field, various physical models are used. In fact, kinetic approach suggested by Hasselman[1] and Zakharov with co-authors (see Refs. 2,3,4) predicts the energetic spectrum shape, as well as its developing in deep water. Just now the study of the evolution of the random surface waves in deep water is performed in the framework of the nonlinear evolution equations like the nonlinear Schrodinger equation (NSE) and its generalization (Dysthe equation) with no using the hypothesis of the momentum closure. Dysthe *et al.*[5] performed the numerical simulation of the NSE with random initial profiles of Gaussian shape. The spectrum broadens symmetrically with time until it reaches a quasi-steady width. If the wave field is initially unstable (Benjamin - Feir instability), its spectrum becomes wide, reducing the instability, and then does not change in the process of the average wave field evolution. The NSE predicts also the generation of the freak waves from the random state[6,7]. The high-order extensions of the NSE on nonlinearity and dispersion (like Dysthe and modified Dysthe equations) change cardinally the process of the random wave evolution. In particular, initial symmetrical spectrum transforms to the asymmetric profile with a steepening of the low-frequency side providing downshift of the spectral peak and an angular widening; equilibrium interval $k^{-2.5}$ is observed (see Ref. 5). The wave process is a Gaussian on average only; its kurtosis oscillates around zero (value for Gaussian distribution), sometimes becomes very high[8]. Large values of the kurtosis correspond to the more energetic tails of the distribution function, providing higher probability of the freak wave occurrence. Also we would like to point out that the extension versions of the NSE (as discrete computed models of the NSE) lead to the chaotic dynamics of the wave field even for regular initial data[9,10].

The main goal of this paper is to study the evolution of the random unidirectional waves in shallow water using the Korteweg – de Vries (KdV) equation. The main difference with the NSE model is an absence of the Benjamin - Feir instability at least for one-peak spectral distribution (recently, Onorato *et al.*[11], demonstrated that two-peak distribution could be unstable). Weakly modulated regular wave packets in the framework of the KdV model split into several groups with different carrier frequency with no formation of large-amplitude waves (see Refs. 12,13). In Ref. 14 it was shown that freak waves can be generated only by the nonlinear – dispersive focusing of the spectral components if the spectrum is relative wide. Analysis of typical realizations of the shallow water waves shows that probability of the high-amplitude waves is less then the Rayleigh prediction[15]. We will consider periodic boundary conditions for the random wave field. The detail analysis of the periodic solutions of the KdV equation expressed through the theta-functions is given in series of papers by Osborne and co-authors (see, for instance, Refs. 16,17,18). The KdV solution in his approach called the nonlinear Fourier method is represented by a linear superposition of the nonlinear oscillatory modes (multi quasi-cnoidal waves) in the associated spectral problem. The number of such modes is not too large as in classic Fourier method, what leads to an effective analysis of the wave field (including random sea state) and selection of the soliton components. Meanwhile, we will use direct Fourier method, taking into account the simple presentation of the results and their comparison with the results of other authors.

## II. BASIC EQUATION

The Korteweg - de Vries equation is considered as a mathematical model for surface gravity weakly nonlinear and weakly dispersive waves

$$\frac{\partial \eta}{\partial t} + \frac{3c\eta}{2h}\frac{\partial \eta}{\partial x} + \frac{ch^2}{6}\frac{\partial^3 \eta}{\partial x^3} = 0, \qquad (1)$$

where $\eta(x,t)$ is a sea surface elevation, $h$ is a water depth assuming being constant, $g$ is a gravitational acceleration, $c = (gh)^{1/2}$ is a maximal speed of long wave propagation, $x$ is a moving frame with long wave speed, $c$, and $t$ is a time. It is the basic equation in the nonlinear theory due to its integrability within the inverse-scattering method (see, for instance, Ref. 19). The main approximations used for derivation (1) are the weakness of nonlinearity, $\varepsilon$, and dispersion, $\mu$

$$\varepsilon = a/h, \qquad \mu = (kh)^2, \qquad (2)$$

where $a$ and $k$ characterize wave amplitude and wave number. The relation between them is defined by the Ursell parameter

$$Ur = \frac{\varepsilon}{\mu} = \frac{A}{k^2 h^3}; \qquad (3)$$

small values of $Ur$ correspond to the almost linear dispersive waves, and large values - to the nonlinear nondispersive waves.

The Korteweg – de Vries equation is the simplest model of the shallow-water waves. Formal estimates show that some observed wind-generated waves in the coastal zone satisfy a weakness of nonlinear and dispersive parameters at specific limitations of the wind parameters and water depth (see Ref. 15). Now more accurate models based on modified

Boussinesq equations are developed[20,21,22]; they can be applied up to full nonlinear and strong dispersive waves.

### III. NUMERICAL MODELING

The numerical integration of the Korteweg -de Vries equation with the periodic boundary conditions $\eta(0,t) = \eta(L,t)$ is based on the pseudospectral method, described in Ref. 23. The random wave field is modeled as the Fourier series

$$\eta(x,t) = \sum_i a_i \cos(k_i x - \omega_i t + \varphi_i), \qquad (4)$$

where $a_i$ is an amplitude of *i*-component with a wave number $k_i$ ($k_i = 2\pi i/L$), and a phase $\varphi_i$ is a random variable in the interval (0-$2\pi$). The spectral amplitudes are constructed from the initial spectrum, *S(k)*, by

$$a_i = \sqrt{2S(k_i)\Delta k}. \qquad (5)$$

Here $\Delta k = 2\pi/L$ is a sampling wave number, equal in our computations to 0.004 m$^{-1}$ (*L* =1570 m). The initial power spectrum, *S(k)* is calculated through the experimental frequency spectrum of wind waves (see Fig. 6 in Ref. 24), measured in the North Sea on the depth *h* = 7 m (Fig. 1), using for simplicity the dispersion relation of the long waves, $S(k)=c \cdot S(\omega)$. The observed wave parameters are significant height, $H_s$ = 2.5 m, and significant period, $T_s$ = 8.4 sec. Thus, the initial spatial spectrum chosen for computing has the significant wavelength, $\lambda_s$ = 68 m, and numerical domain contains approximately 30 individual waves. We will characterize the energetic spatial spectrum by the significant amplitude, defined as $A_s = 2\sigma$, where the variance, $\sigma^2$ is determined by $\sigma^2 = 2\int S(k)dk$. For the calculated initial spectrum, $A_s$ =1.25 m; thus, the parameters of the nonlinearity and dispersion are, $\varepsilon$ = 0.18 and $\mu$ = 0.42, so the Ursell parameter, *Ur*, is 0.43. Two series of experiments are conducted to investigate the influence of nonlinearity and dispersion independently.

### A. Influence of nonlinearity

To demonstrate the influence of nonlinearity, several spectra of the same shape with different significant wave amplitudes (and the same wavelength) are considered. The first spectrum is an observed one presented in Fig. 1. The rest spectra are obtained from the observed spectrum by its multiplying on constant numbers. All the parameters of experiments are summarized in Table 1.

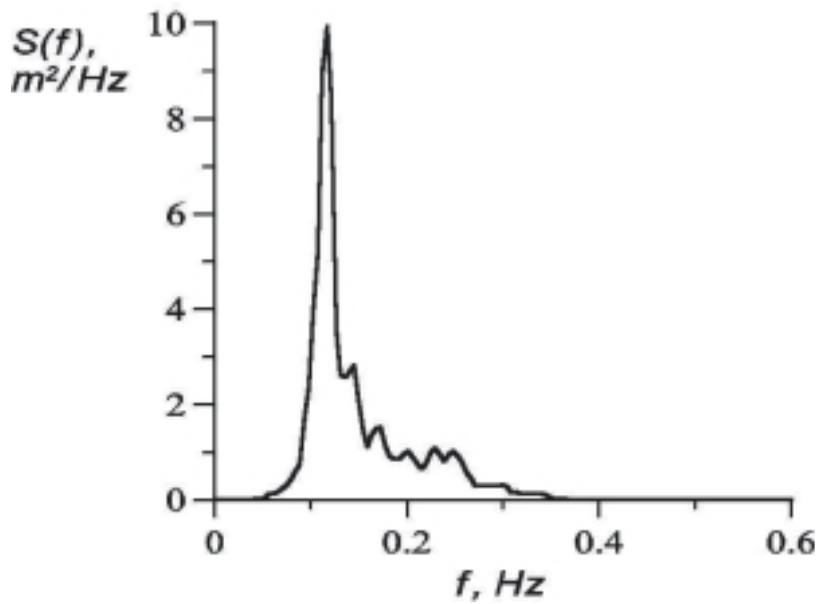

**Figure 1.** Observed power spectra of wind-generated waves in the North Sea
($h$=7 m, $H_s$ =2.5 m, $T_s$=8.4 sec)

Table I. Parameters of experiments

| $A_s$, m | $\varepsilon$ | Ur | $A_s$, m | $\varepsilon$ | Ur |
|---|---|---|---|---|---|
| 0.42 | 0.06 | 0.14 | 1.90 | 0.27 | 0.64 |
| 0.89 | 0.13 | 0.30 | 2.10 | 0.30 | 0.72 |
| 1.25 | 0.18 | 0.43 | 2.94 | 0.42 | 1.00 |

The computing is done for relative large times (10 min in real time, or about 100 fundamental wave periods). The number of realizations used for analysis is 500, what corresponds to the wave record with about 15000 individual waves. The time evolution of the spectrum and distribution functions (water elevation and crest amplitude) is determined. Also, the skewness, $m_3$, kurtosis, $m_4$, of the wave field are calculated (remain, that the first two moments: mean level and variance, are the integrals of the KdV equation, and they are constant in the process of wave evolution). For skewness and kurtosis the standard definitions are used:

$$m_3 = \frac{M_3}{\sigma^3}, \qquad m_4 = \frac{M_4}{\sigma^4} - 3, \qquad (6)$$

where $M_3$ and $M_4$ are the third and the fourth statistical moments of wave field consequently. As it is known, the skewness is a statistical measure of the vertical asymmetry of the wave field, and its sign defines the ratio of crests to troughs. If it is positive, then the crests are bigger then the troughs. The kurtosis represents a degree of the peakdness in the distribution and defines the contribution of the big waves. If it is positive, then the contribution of the big waves is rather significant.

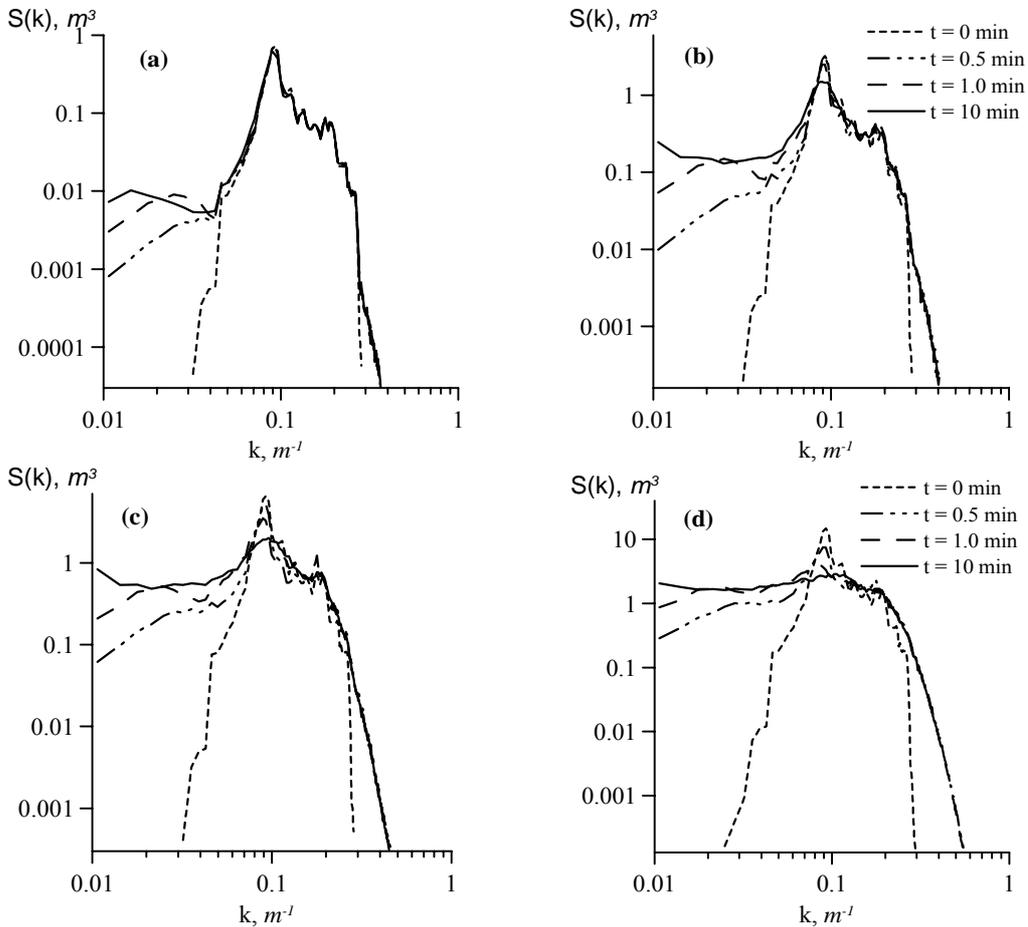

**Figure 2.** Temporal evolution of spectra $S(k,t)$ for different values of nonlinear parameter, (a) $\varepsilon = 0.06$, (b) $\varepsilon = 0.13$, (c) $\varepsilon = 0.18$, (d) $\varepsilon = 0.42$

Figures 2(a)-2(d) demonstrate the spectra evolution for different nonlinear parameters during 10 minutes (approximately 10 characteristic nonlinear times, $T_s/\varepsilon$). It could be observed that for small nonlinearity parameter [Fig. 2(a)] spectrum is weakly varied in time,

and for bigger ε [Fig. 2(d)] the spectrum broadens during the first minutes (characteristic nonlinear time, $T_s/\varepsilon$) until it reaches the stationary state, and the energy of the long waves is distributed almost uniformly. The flatness of spectrum is bigger for spectrum (d), when the nonlinear effects are more significant. For the larger $k$ (0.1 < k < 0.2), the spectrum can be approximated by power asymptotic, $k^{-\alpha}$, where the slope of the spectrum, $\alpha$ is decreased from 4.2 to 0.4 with increasing of wave amplitude. It is important to mention that the spectrum downshifting in deep water can be obtained only in extended version of the NSE[5] included asymmetry of the wave field. The KdV model is initially asymmetric due to quadratic nonlinearity, and the asymmetry of the regular wave group is immediately obtained in the process of the wave evolution[12,25].

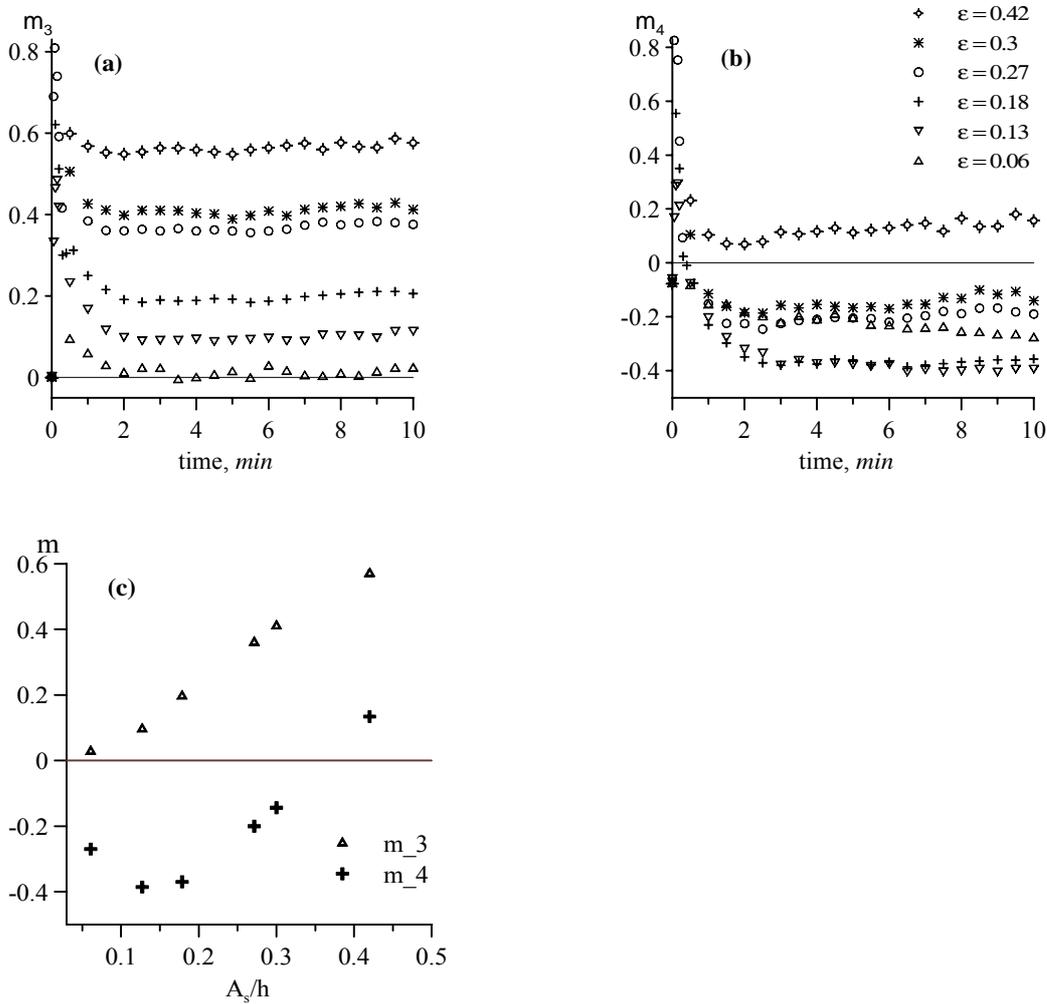

**Figure 3.** Temporal evolution of the skewness, $m_3$, (a) and kurtosis, $m_4$, (b) for different values of nonlinear parameter and (c) - relation of the moments asymptotic values with nonlinear parameter

In Figs. 3(a) and 3(b) the temporal evolution of skewness and kurtosis for different ε is shown. The initial wave field taken from the observations is not truly Gaussian process. After short transition period, both moments of the wave field tend to the almost constant values during few characteristic nonlinearity times. For all the conditions, the skewness is positive, and it means that the positive waves (crests) have larger amplitudes then the negative waves (troughs). The asymptotic value of skewness increases with increasing of the wave amplitude (nonlinearity), from 0.1 to 0.6; and, therefore, positive waves are more visible in the nonlinear wave field then the negative waves. It corresponds to the well-known opinion that solitons should play an important role in the dynamics of the wave field for large amplitudes. The kurtosis tends to the negative asymptotic value for $\varepsilon < 0.4$; therefore, the probability of the large-height (freak) wave occurrence should be less than it is predicted for the Gaussian processes. It is interesting to note that the module of the kurtosis decreases (from 0.4 to 0.14) with increasing of the wave amplitude (nonlinearity), and, therefore, the occurrence of the freak wave becomes more often, if the background wave field is more energetic. In the case of strong nonlinearity, the kurtosis asymptotic value exceeds zero level, what indicates a high probability of the large wave appearance. For deep water, as it was already shown by Onorato *et al.*[8] and Tanaka[26], the calculated values of kurtosis oscillate around zero with some exceeding for nonlinear random wave process, what indicates that probability of large waves is bigger if nonlinearity is taken into account (see Ref. 8), and kurtosis takes on a positive values.

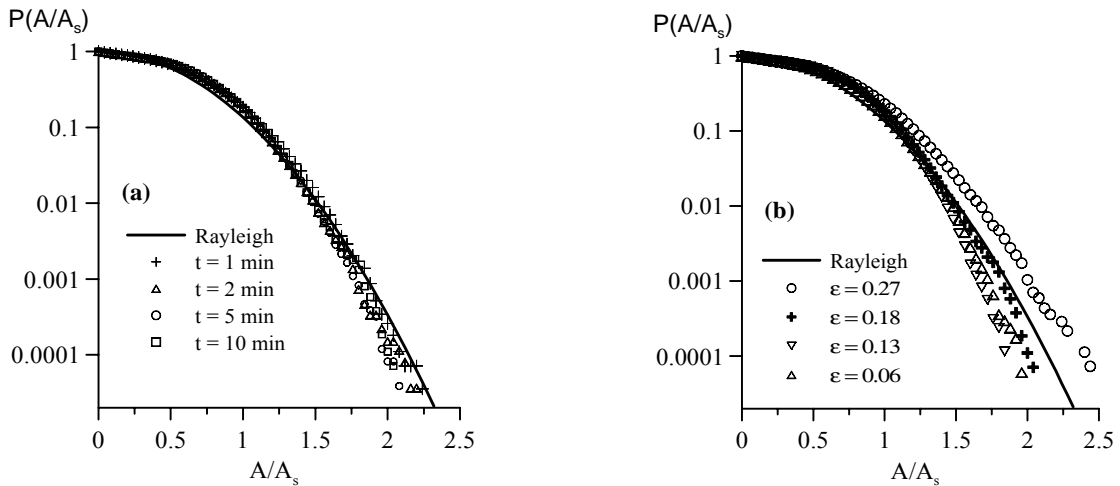

**Figure 4.** Crest amplitude distribution: (a) time evolution for $\varepsilon = 0.18$, and (b) asymptotic distribution for different nonlinear parameters. Solid line corresponds to the Rayleigh distribution of the narrow-band Gaussian process

Distribution of the wave amplitudes, calculated as maximum between two zeros, is demonstrated in Fig. 4. The results are compared with the theoretical Rayleigh distribution of the amplitudes of the narrow-band Gaussian process. For experiments with the observed spectrum ($\varepsilon = 0.18$), the probability of small amplitudes ($A < 1.2A_s$) is bigger then the Rayleigh distribution, meanwhile in the range of high amplitudes ($A > 1.5A_s$) the distribution lays below theoretical curve. If the wave field is energetic ($\varepsilon = 0.27$), the asymptotic distribution exceeds the Rayleigh distribution [Fig. 4(b)], and the probability of highest crests appearance increases. Quality, the shape of the amplitude distribution function is not in contradiction with the behavior of the skewness and kurtosis (Fig. 3); the first one shows the positive waves have larger amplitudes than negative waves, meanwhile the second one indicates significant contribution of the small waves in total distribution. It is proofed with Fig. 5, where distribution of trough amplitudes is shown. It is seen, that amplitudes of troughs become smaller with the increasing of nonlinearity and lay below Rayleigh distribution, while the wave crests grow with $\varepsilon$.

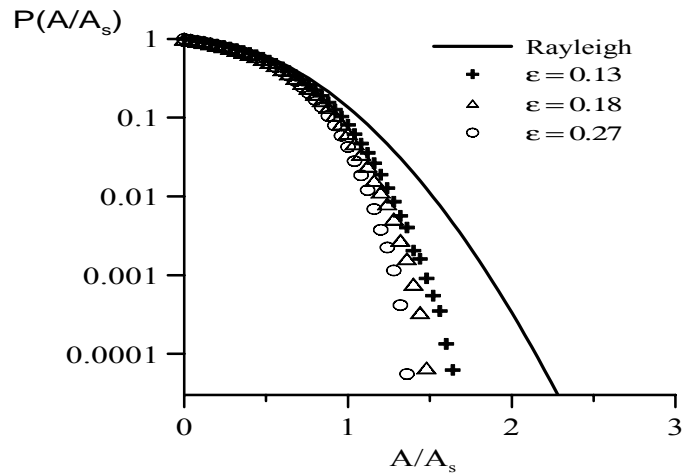

**Figure 5.** Asymptotic distribution of the trough amplitudes for different nonlinear parameters. Solid line corresponds to the Rayleigh distribution of the narrow-band Gaussian process

### B. Influence of dispersion

The second series of experiments represents the influence of the dispersion parameter. For this purpose, the spectra with the nonlinear parameters $\varepsilon = 0.13$ used above, as whole are shifted down and up the wave number axis without change of the significant wave amplitude. As a result, the significant wavelength is varied from 25 to 110 m. The initial profiles of power spectra are presented in Fig. 6(a) with the corresponding dispersion parameter, $\mu$, set to 0.16, 0.24, 1.04 and 2.95, and Ursell numbers equal to 0.82, 0.53, 0.12, 0.04. The computing is done for 10 min. All the estimations are made with the averaging over 500 realizations.

Figure 6(b), represented the spectra shape at time $t = 10$ min, demonstrates the widening of the spectrum till the almost uniform energy distribution for long wave side ($k < 0.1\ m^{-1}$) in the case of small dispersion parameters (the initial energetic peak wave number is less then $k_m =$

0.1 m$^{-1}$), what is adjusted with the results discussed in section III.A. In the case of big dispersion ($k_m > 0.1$ m$^{-1}$) the spectrum cannot broaden uniformly in the low frequencies due to the strong dispersion effect, but it also develops till some steady level is reached.

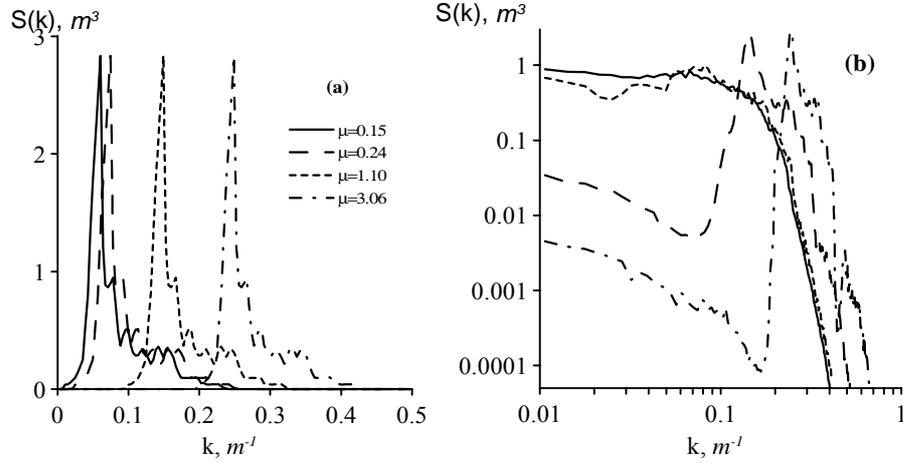

**Figure 6.** Initial spectra (a) and (b) - asymptotic steady spectra (t = 10 min) for different values of dispersion parameter

In Fig. 7(a) and 7(b) the temporal evolution of skewness and kurtosis is presented, and Fig. 8 demonstrates the dependence of the asymptotic values of $m_3$ and $m_4$ on the dispersion parameter for the case $\varepsilon = 0.13$. As in previous case (Sec. III.A) the tendency of positive skewness is observed. With increasing of dispersion (wave becomes more linear), the skewness monotonic tends to zero (linear wave limit). Another behavior is observed for kurtosis characteristic: for large values of dispersion, kurtosis does not exceed its initial value, and for smaller dispersion there is high exceeding of initial level during a few characteristics nonlinear time and then it tends to a quasi-state level, which value grows with dispersion decreasing. It correlates with the same dependence from the nonlinear parameter [Fig. 3(c)].

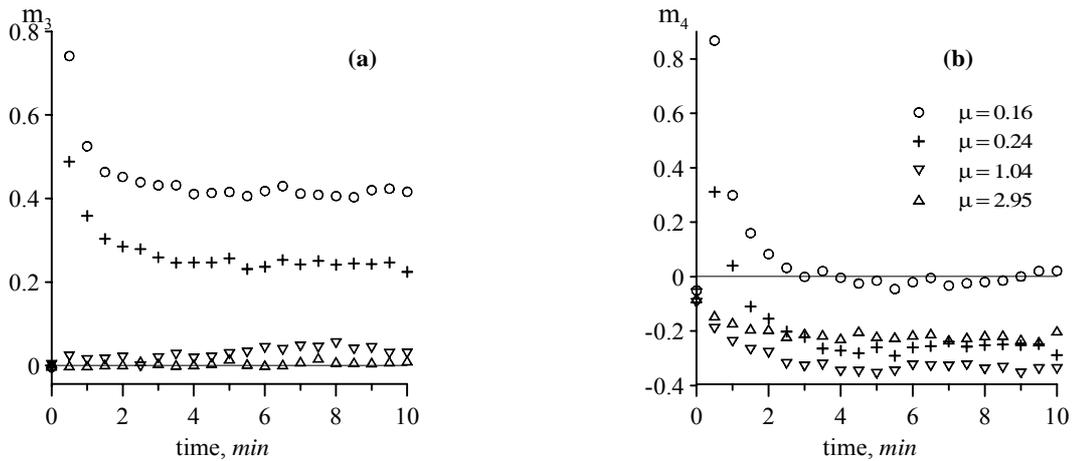

**Figure 7.** Temporal evolution of skewness (a) and kurtosis (b) for different $\mu$

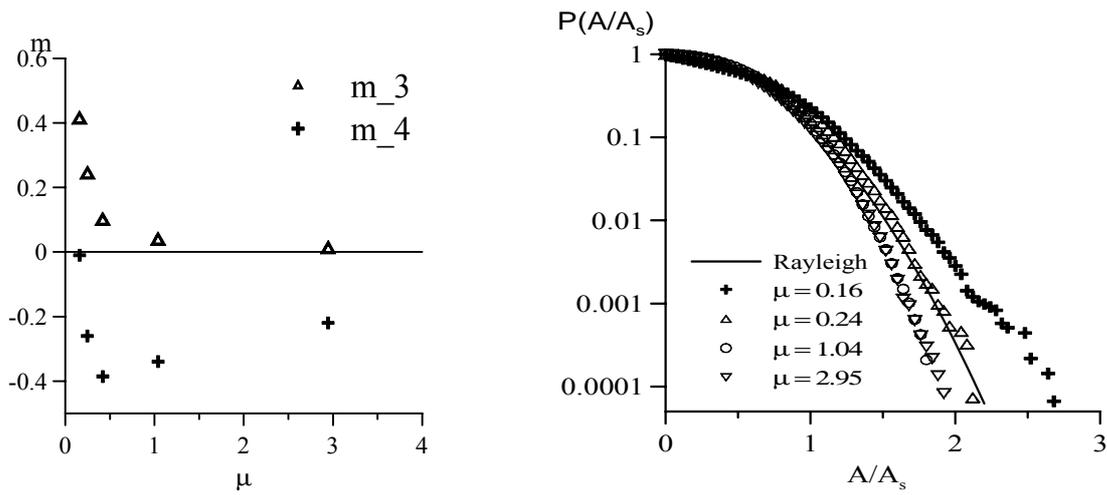

**Figure 8.** Dependence of skewness and kurtosis asymptotic values on the dispersion parameter

**Figure 9.** Asymptotic crest distribution for different dispersion parameters. Solid line corresponds to the Rayleigh distribution of the narrow-band Gaussian process

The distribution of wave crest amplitude for different $\mu$ at time $t = 10$ min is calculated (Fig. 9). If dispersion is significant then amplitude probability lays below Rayleigh statistics, and for weak dispersion (amplitude keeps to be constant) the nonlinear effects become prevail, what gives an increasing of the amplitude probability.

## IV. CONCLUSION

Let us summarize the characteristics of the random wave field as a function of the Ursell parameter. In Fig. 10 the generalized dependence of the skewness and kurtosis asymptotic values on the Ursell parameter for the different amplitudes and wavelength according to section III.A and III.B are presented. As it is seen, if the parameter Ur is rather small, what corresponds to the linear dispersive waves, the asymmetry of the wave field is also small and kurtosis is close to the initial value. The increasing of the nonlinear effects (big Ur number) leads to the increasing of asymmetry in the wave field with bigger crests amplitudes then troughs (growing of positive skewness). The behavior of the kurtosis is non-monotonic, and for large $Ur$ (~1.0) it becomes positive and contribution of large-height waves becomes significant in comparison with the small-height waves.

In Fig. 11 the amplitude crest distribution in respect to Ursell number is presented to summarize the results, obtained in Sec. III.A and III.B. It demonstrates and confirms the influence of nonlinear effects on the large amplitudes appearance: its probability increases if nonlinearity is significant, and for small $Ur$, amplitude distribution does not exceed the initial distribution and probability of the large waves is less then theoretical one.

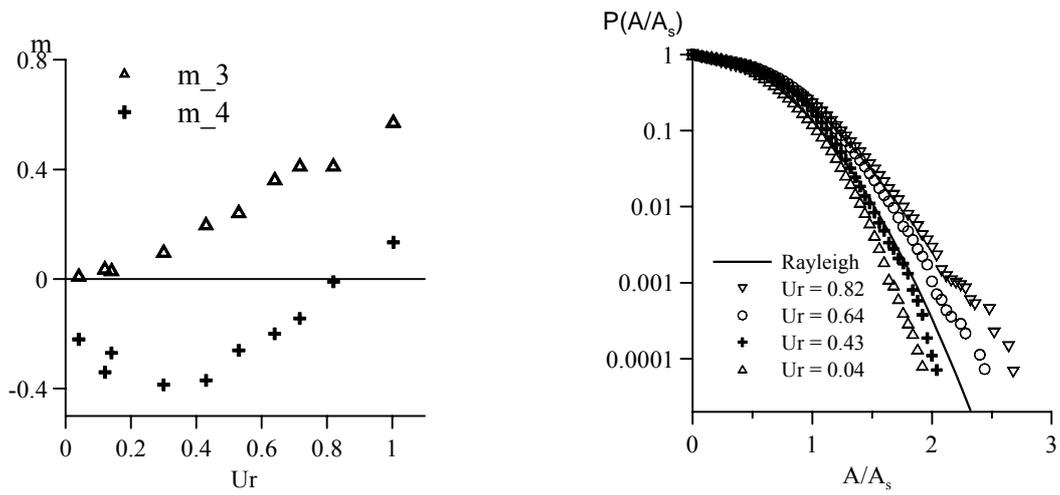

**Figure 10.** The relation of the asymptotic values of the moments with parameter Ur

**Figure 11.** Asymptotic crest distribution for different *Ur* numbers. Solid line corresponds to the Rayleigh distribution of the narrow-band Gaussian process

The temporal spectra evolution shows the spectrum broadening till the almost uniform energy distribution in the range of long waves ($k < 0.1\ m^{-1}$) for large *Ur* numbers ($Ur > 0.43$). For the shortest waves ($0.1 < k < 0.15$) the slope of the spectra, α, in the power-low asymptotic is not a constant; it is shown in Fig. 12. The slope decreases with the increasing of the Ursell values and spectrum tends to the $k^{-1}$ asymptotic in our experiments.

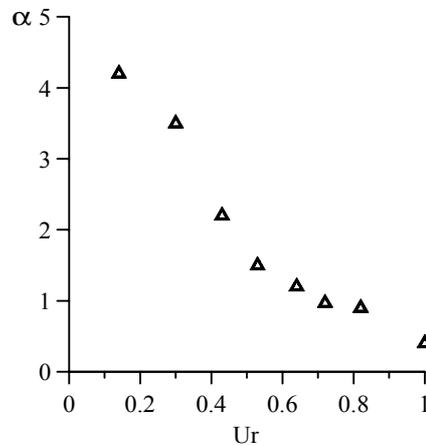

**Figure 12.** Spectral asymptotic ($0.09 < k < 0.15$) at time $t = 10$ min for different *Ur* numbers

The obtained results allowed to study the dynamics of unidirectional random waves in shallow water within the KdV model. Numerical experiments demonstrated dependence of wave statistics on the ratio of nonlinear effect to dispersion, what leads to the growing of the asymmetry in wave field with time and increasing of the contribution of the large waves in total distribution. Numerical results showed dependence of the wave amplitude probability on the *Ur* parameters. It demonstrated that for *Ur>0.5* the probability of large amplitudes (more then $2A_s$) exceeds the Rayleigh distribution.

## ACKNOWLEDGMENTS

The research is supported by RFBR grants (02-05-65107, 03-05-06075) and INTAS grant (01-2156).

## References

[1] K. Hasselmann, "On the nonlinear energy transfer in a gravity wave spectrum," J Fluid Mech. **12**, 481 (1962); **15**, 273 (1963).

[2] V.E. Zakharov, and N. Filonenko, "The energy spectrum for stochastic oscillation of a fluid's surface," Doklady Akad. Nauk USSR **170**, 1292 (1966).

[3] V.E. Zakharov, and M.M. Zaslavsky, "The kinetic equation and Kolmogorov spectra in the weak turbulence theory of wind waves," Izv., Atm. Ocean Phys. **18**, 747 (1982).

[4] V.E. Zakharov, "Statistical theory of gravity and capillary waves on the surface of finite-depth fluid," European J. Mech., B/Fluids **18**, 327 (1999).

[5] K.B. Dysthe, K. Trulsen, H.E. Krogstad, and H. Socquet-Juglard, "Evolution of a narrow-band spectrum of random surface gravity waves," J.Fluid Mech. **478**, 1 (2003).

[6] M. Onorato, A.R. Osborne, M. Serio, and S. Bertone, "Freak wave in random oceanic sea states," Phys. Rev. Lett. **86**, 5831 (2001).

[7] A. Slunyaev, C. Kharif, E. Pelinovsky, and T. Talipova, "Nonlinear wave focusing on water of finite depth," Physica D **173**, 77 (2002).

[8] M. Onorato, A.R. Osborne, and M. Serio, "Extreme wave events in directional, random oceanic sea states," Phys. Fluids **14**, 25 (2002).

[9] M.I. Ablowitz, J. Hammack, D. Henderson, and C.M. Schober, "Modulated periodic Stokes waves in deep water," Phys. Rev. Lett. **84**, 887 (2000); "Long-time dynamics of the modulational instability of deep water waves," Physica D **152-153**, 416 – 433 (2001).

[10] A. Calini. and C.M. Schober, "Homoclinic chaos increases the likelihood of rogue wave formation," Phys. Lett. A **298**, 335 (2002).

[11] M. Onorato, D. Ambrosi, A.R. Osborne, and M. Serio, "Instability of two interacting, quasi-monochromatic waves in shallow water," Phys. Fluids (2003) (submitted).

[12] E. Kit, L. Shemer, E. Pelinovsky, T. Talipova, O. Eitan, and H. Jiao, J. "Nonlinear wave group evolution in shallow water," J. Waterway, Port, Costal, Ocean Eng. **126**, 221 (2000).

[13] E.van Groessen, and J.H.Westhuis, Mathematics and Computers in Simulation **59**, 341 (2002).

[14] E. Pelinovsky, T. Talipova, and C. Kharif, "Nonlinear dispersive mechanism of the freak wave formation in shallow water," Physica D **147**, 83 (2000).

[15] A. Kokorina, and E. Pelinovsky, "Applicability of the Korteweg-de Vries equation for description of the statistics of freak waves," J. Korean Society Coastal and Ocean Engineers, **14**, 308 (2002).

[16] A.R. Osborne, E. Segre and G. Boffetta, "Soliton basis states in shallow-water ocean surface waves," Phys. Rev. Lett. **67**, 592 (1991).

[17] A.R. Osborne, "Numerical construction of nonlinear wave-train solutions of the periodic Korteweg – de Vries equation," Phys. Rev. E **48,** 296 (1993); "Solitons in the periodic Korteweg – de Vries equation, the Θ-function representation, and the analysis of nonlinear, stochastic wave trains," Phys. Rev. E **52**, 1105 (1995); "Behavior of solitons in random-function solutions of the periodic Korteweg – de Vries equation," Phys. Rev. Lett. **71**, 3115 (1993).

[18] A.R. Osborne, M. Serio, L. Bergamasco, and L. Cavaleri, "Solitons, cnoidal waves and nonlinear interactions in shallow-water ocean surface waves," Physica D **123**, 64 (1998).

[19] P.G. Drazin, and R.S. Johnson, *Solitons: An introduction* (Cambridge Univ. Press, 1993).

[20] G. Wei, J.T. Kirby, S.T. Grilli, and R. Subramanya, "A fully nonlinear Boussinesq model for surface waves," J. Fluid Mech. **294**, 71 (1995).

[21] P.A. Madsen, H.B. Bingham, and H. Liu, "A new Boussinesq method for fully nonlinear waves from shallow to deep water," J. Fluid Mech. **462**, 1 (2002)

[22] P.A. Madsen, H.B. Bingham, and H.A. Schaffer, "Boussinesq-type formulations for fully nonlinear and extremely dispersive water waves: derivation and analysis," Proc. R. Soc. London A **459**,1075 (2003).

[23] B. Fronberg, *A Practical Guide to Pseudospectral Methods* (Cambridge Univ. Press, 1998).

[24] R.C. Ris, L.H. Holthuijsen, and N. Booij, "A third –generation wave model for coastal regions," J. Geophys. Research **C4**, 7667 (1999).

[25] R. Grimshaw, D. Pelinovsky, E. Pelinovsky, and T. Talipova, "Wave group dynamics in weakly nonlinear long-wave models," Physica D **159**, 35 (2001)

[26] M. Tanaka, "A method of studying of nonlinear random field," Fluid Dyn. Res., **28**, 41 (2001).